\documentclass[conference]{IEEEtran}
\IEEEoverridecommandlockouts
\usepackage{graphics}
\usepackage{cite}
\usepackage{amsmath,amssymb,amsfonts}
\usepackage{amsmath}
\usepackage{algorithmic}
\usepackage{graphicx}
\usepackage{textcomp}
\usepackage{xcolor}
\usepackage{subfigure}
\usepackage{algorithm}

\usepackage{multirow}
\usepackage{multicol}
\usepackage{array,multirow}
\usepackage{breqn}

\usepackage[T1]{fontenc}
\usepackage{mathtools}   

\DeclarePairedDelimiter{\norm}{\lVert}{\rVert}

\def\BibTeX{{\rm B\kern-.05em{\sc i\kern-.025em b}\kern-.08em
    T\kern-.1667em\lower.7ex\hbox{E}\kern-.125emX}}
    
\begin{document}

\title{Real-Time  Power System Dynamic Simulation using Windowing based Waveform Relaxation Method   
\thanks{This work is supported in part by the U.S. Department of
Energy’s Office of Energy Efficiency and Renewable Energy (EERE) under the Solar Energy Technologies Office Award Number DE-EE0008774, National Science Foundation grant ECCS-1810174, and National Science Foundation grant ECCS-2001732. Corresponding Author: M Al Mamun, Florida International University, Email: mmamu011@fiu.edu}
}   

 \author{%
\IEEEauthorblockN{M Al Mamun$^\Diamond$, Sumit Paudyal$^\Diamond$, and  Sukumar Kamalasadan$^\oplus$}
\IEEEauthorblockA{
$^\Diamond$Florida International University, USA; $^\oplus$University of North Carolina at Charlotte, USA \\
Emails: mmamu011@fiu.edu,  spaudyal@fiu.edu, skamalas@uncc.edu } \vspace{-25pt}
}
\maketitle

\begin{abstract}
Power system dynamic modeling involves nonlinear differential and algebraic equations (DAEs). Solving DAEs for large power grid networks by direct implicit numerical methods could be inefficient in terms of solution time; thus, such methods are not preferred when real-time or faster than real-time performance is sought. Hence, this paper revisits  Waveform Relaxation (WR) algorithm, as a distributed computational technique to solve power system dynamic simulations.
{\color{black} Case studies performed on the IEEE NE 10-generator 39-bus system demonstrate that, for a certain simulation time window, the solve time for WR method is larger than the length of the simulation window; thus, WR lacks the performance needed for real-time simulators, even for a small power network. To achieve real-time performance, then a Windowing technique is applied on top of the WR, for which the solve time was obtained less than the length of a simulation window, that shows the effectiveness of the proposed method for real-time dynamic simulation of power systems.} 
\end{abstract}

\vspace{7pt}

\section{Introduction}
Dynamic simulation is crucial for designing reliable operation, control, and protection of electrical power systems. For the complete dynamic behavior, the mathematical modeling of a power system, in its original form, involves nonlinear differential and algebraic equations (DAEs). 
Direct integration (DI) is the commonly used method in solving power system dynamic models, where the numerical techniques such as  Backward Euler, Trapezoidal, Runge Kutta \cite{hairer2006numerical}, Block Backward Difference formula (BDF) \cite{jiaxiang1995numerical} are used to discretize the system of DAEs. This discretization results in a set of nonlinear equations, whose solution process involves iteratively solving a set of linear equations in the form $\mathbf{[A]\,[x]= [b]}$. Generally, the  DI method is implemented sequentially in a single processing unit. However, for real-time needs, the sequential approach of DI for solving large power networks could be inefficient in terms of computational time. Thus, the distributed or parallel approach of computation in the DI method to solve the problem using a multi-processor environment is desired \cite{burrage1993parallel}. For parallel realization,   $\mathbf{[A]}$   can be decomposed to block bordered diagonal form  \cite{happ1979future} or  L and U triangular
matrices \cite{4111840}. Also, Runge-Kutta method in \cite{van1990parallel, liu2011parallel} and trapezoidal method  in \cite{alvarado1979parallel} are modified for parallel realization. 
However, DI methods discretize the state variables at a uniform step size. Since state variables  change at different  rates, the DI method uses the smallest possible time step to effectively capture the dynamic behavior of the fastest-changing variable, which increases the
computational burden.     

Waveform Relaxation (WR) is an iterative approach of numerical integration to solve the system of ODEs/DAEs. The WR algorithm was initially introduced in VLSI circuits as a numerical integration method to solve ODEs \cite{white2012relaxation}. Power system models are very similar to VLSI circuits in terms of transient behavior and, and hence WR algorithms have been successfully applied to power system problems \cite{crow1990parallel}. The seminal works on WR applied to power system stability analysis are in \cite{crow1990parallel, liu2015two}.  In the WR method, the physical system is first divided into multiple subsystems and the variables associated with each subsystem are solved independently \cite{crow1994waveform}. 
Gauss Jacobi version of WR is parallel in nature \cite{sand1998jacobi}. Thus, it is possible to have a distributed or parallel computation where each subsystem is solved independently by one individual processing unit. As the computational burden can be distributed over the available processors, WR could be more efficient than that of the DI method for larger systems. Moreover, if the system splitting could separate the strongly coupled variables and weakly coupled variables and divide them into subsystems accordingly, it could be possible to discretize the state variables at different step sizes \cite{crow1994waveform}, making the computation more efficient. However, a good partitioning algorithm is crucial for faster convergence of the WR algorithm. Arbitrary splitting of the network may lead to slower convergence and, even, diverged solution \cite{lelarasmee1982waveform}.

WR generally requires a larger number of iterations to converge even if the system is properly partitioned; thus, this method could still be slower compared to the DI method. To overcome this drawback, the entire simulation duration can be split into several smaller and identical windows. Then, the WR algorithm is applied to each of the time windows individually and separately, which could lead to time parallel operation on top of the conventional WR and make the overall computation faster than the DI method \cite{crow1990parallel}. This method is called WR with Windowing (WRW) technique. WRW was first introduced in \cite{sangiovanni1985waveform} and, after that, it was greatly used for parallel computation over parallel processors \cite{smart1988waveform, jalili2009instantaneous}.  

The main contributions of this work are comparative analysis of WR and WRW with that of the DI method for power system dynamic simulations, and performance analysis of WR and WRW methods varying the number of subsystems and simulation periods. For the rest of the paper, section II describes the detailed concept of WR and WRW algorithms, section III provides the dynamic modeling of power systems, numerical simulation results are shown in section IV, and section V concludes the paper with future work directions.

\section{Algorithms}
\subsection{Waveform Relaxation Algorithm}
As WR is an iterative algorithm, it requires an initial guess for all the system variables for the entire duration of the simulation. Then, the system variables are grouped into multiple subsystems based on a certain splitting criterion. Then, each subsystem is solved independently using solutions for the variables from other subsystems as inputs. After each iteration of WR, solutions are exchanged among the subsystems. Then, the next iteration of WR starts with improved solutions and it continues until all of the variables have converged solutions. In Gauss Jacobi WR, while solving a specific subsystem variables, solutions from previous WR iteration are used as inputs for the variables from other subsystems. On the other hand, Gauss-Seidel uses the latest solutions from other subsystems, which could be the solution from the same WR iteration or the previous WR iteration depending on the order subsystems are solved \cite{smart1988waveform}. 

Consider a dynamic system which can be described by the following DAEs,  
\begin{equation}
\begin{aligned}
    \mathbf{\dot x = f(x,y)}\\ 
    \mathbf{0 = g(x,y)},
\label{eq1}
\end{aligned}
\end{equation}
where $\mathbf{x}$ is a vector of state variables and $\mathbf{y}$ is a vector of algebraic variables. For this system of DAE, the WR partitions the system into $p$ number of subsystems and, then, discretizes the subsystem equations as below, using any of the implicit integration methods \cite{zecevic1999partitioning}.
\begin{equation}
\begin{aligned}
    \dot x_i = f_i(x_1,..,x_i,..,x_p, y_1,..,y_i,..y_p), \forall i=1,..,p\\ 
    0 = g_i(x_1,..,x_i,..,x_p, y_1,..,y_i,..y_p), \forall i=1,..,p,
\label{eq2}
\end{aligned}
\end{equation}
where $x_i$ and $y_i$  are vectors of state variables and algebraic variables of the $i^{th}$ subsystem, respectively.  Now, each of the subsystems can be solved iteratively using Gauss Jacobi WR shown below for $k^{th}$ iteration. The detailed procedure is shown in Algorithm~\ref{alg:algorithm1}, where it is evident that the
subsystems can be solved in parallel in Gauss Jacobi WR algorithm. 
\vspace{-2pt}
\begin{equation}
\begin{aligned}
    \dot x_i^{k} = f_i(x_1^{k-1},..,x_i^k,..,x_p^{k-1}, y_1^{k-1},..,y_i^k,..y_p^{k-1})\\ 
    0 = g_i(x_1^{k-1},..,x_i^k,..,x_p^{k-1}, y_1^{k-1},..,y_i^k,..y_p^{k-1}).
\label{eq3}
\end{aligned}
\end{equation}
For Gauss Seidel WR, the following iterative equations are used.
\vspace{-5pt}
\begin{equation}
\begin{aligned}
    \dot x_i^{k} = f_i(x_1^{k},..,x_i^{k},..,x_p^{k-1}, y_1^{k},..,y_i^{k},..y_p^{k-1})\\ 
    0 = g_i(x_1^{k},..,x_i^{k},..,x_p^{k-1}, y_1^{k},..,y_i^{k},..y_p^{k-1}).
\label{eq4}
\end{aligned}
\end{equation}

\begin{algorithm} 
\small
\begin{algorithmic}
  \STATE $k \leftarrow 0$
  \STATE Initialize $\textbf{x}^k(t)$ for $t \in  [0,T]$, such that $\textbf{x}^{k}(0)=\textbf{x}_0$.
  \STATE Initialize $\textbf{y}^k(t)$ for $t\in [0,T]$, such that $\textbf{y}^{k}(0)=\textbf{y}_0$.
  \REPEAT
    \STATE $k \leftarrow k+1$
    \FOR {($i = 1$ to $p$)}
    \STATE $\dot x_i^{k} = f_i(x_1^{k-1},..,x_i^k,..,x_p^{k-1},y_1^{k-1},..,y_i^k,..y_p^{k-1});$
    \STATE $x_i^k(0)=x_0$.
    \STATE $0 = g_i(x_1^{k-1},..,x_i^k,..,x_p^{k-1},y_1^{k-1},..,y_i^k,..y_p^{k-1});$
    \STATE $y_i^k(0)=y_0$.
    \ENDFOR
  \UNTIL{($\norm{\textbf{x}^{k}-\textbf{x}^{k-1}} \leq \varepsilon$ and $\norm{\textbf{y}^{k}-\textbf{y}^{k-1}} \leq \varepsilon$);  where $\varepsilon$ is a small and positive value.}
\end{algorithmic}
\caption{Gauss Jacobi WR Algorithm}
\label{alg:algorithm1}
\end{algorithm}


\subsection{Waveform Relaxation with Windowing Technique}
 WRW divides the entire simulation time interval into several sub-intervals, which are called Windows.  Regular WR algorithm is applied on each Windows sequentially (or in time parallel). In sequential solving approach, computation in a window begins by taking the  solution of the last time step from the previous window as the initial values.  This process is repeated until the last Window is solved \cite{jalili2010acceleration}. The complete procedure is summarized  in Algorithm~\ref{alg:algorithm2}, where  $w_{max}$ is the maximum number of windows and $T_{win}$ is the length of each window. 
\vspace{-1pt}
\begin{algorithm} 
\small
\begin{algorithmic}
  \FOR{($w = 1$ to $w_{max}$)}
    \IF{($w==1$)}
      \STATE $k \leftarrow 0$
      \STATE Initialize $\textbf{x}^k(t)$ for $t \in  [0,T_{win}]$, such that $x^{k}(0)=x_0$.
      \STATE Initialize $\textbf{y}^k(t)$ for $t \in  [0,T_{win}]$. such that $y^{k}(0)=y_0$.
      
      \REPEAT
      \STATE $k \leftarrow k+1$
      \FOR {($i = 1$ to $p$)}
      \STATE $\dot x_i^{k} = f_i(x_1^{k-1},..,x_i^k,..,x_p^{k-1},y_1^{k-1},..,y_i^k,..y_p^{k-1});$
      \STATE $x_i^k(0)=x_0$.
      \STATE $0 = g_i(x_1^{k-1},..,x_i^k,..,x_p^{k-1},y_1^{k-1},..,y_i^k,..y_p^{k-1});$
      \STATE $y_i^k(0)=y_0$.
      \ENDFOR
    \UNTIL{($\norm{\textbf{x}^{k}-\textbf{x}^{k-1}} \leq \varepsilon$ and $\norm{\textbf{y}^{k}-\textbf{y}^{k-1}} \leq \varepsilon$).}
     
    \ELSE
      \STATE $k \leftarrow 0$
      \STATE Initialize $\textbf{x}^k(t)$ for $t \in  [(w-1)T_{win}, wT_{win}]$, such that $x^{k}((w-1)T_{win})=x((w-1)T_{win})$.
      \STATE Initialize $\textbf{y}^k(t)$ for $t \in  [(w-1)T_{win}, wT_{win}]$, such that $y^{k}((w-1)T_{win})=y((w-1)T_{win})$.
      
      \REPEAT
      \STATE $k \leftarrow k+1$
      \FOR {($i = 1$ to $p$)}
      \STATE $\dot x_i^{k} = f_i(x_1^{k-1},...,x_i^k,...,x_p^{k-1},y_1^{k-1},...,y_i^k,...y_p^{k-1});$
      \STATE $x_i^k((w-1)T_{win})=x_{i,(w-1)T_{win}}$.
      \STATE $0 = g_i(x_1^{k-1},...,x_i^k,...,x_p^{k-1},y_1^{k-1},...,y_i^k,...y_p^{k-1});$
      \STATE $y_i^k((w-1)T_{win})=y_{i,(w-1)T_{win}}$.
      \ENDFOR
      \UNTIL{($\norm{\textbf{x}^{k}-\textbf{x}^{k-1}} \leq \varepsilon$ and $\norm{\textbf{y}^{k}-\textbf{y}^{k-1}} \leq \varepsilon$).}
    \ENDIF 
  \ENDFOR  
\end{algorithmic}
\caption{Gauss Jacobi WR with Windowing Technique}
\label{alg:algorithm2}
\end{algorithm}


\section{Power System Dynamic Model}
The dynamic model of power system involves generators' dynamic state variables such as rotor angle and angular frequency, and network algebraic power flow variables which are bus voltage magnitude, voltage angle, real and reactive power generations. In this work, classical model of generators is considered, which is commonly used to study the rotor angle dynamics \cite{cadeau2011coupling}. The  DAEs of power system is  discretized (with step size of $h$) using the Backward Euler method, which results in a system of following time-coupled nonlinear equations.  
\vspace{-3pt}  
\begin{align}
&\delta_m(t+1)=\delta_m(t) + h[\omega_{m}(t+1)-\omega_s] \\
&\omega_m(t+1)=\omega_m(t)+h\left[\frac{1}{M_m}(Pm_m-Pe_m(t+1)\right] \\
\label{dynamic_dis}
& Pe_r(t+1)=P_{Lr}+V_r(t+1) \sum_{j=1}^{N} V_j(t+1) \nonumber \\ 
& \hspace{1.5cm} [B_{rj} sin(\delta_{rj}(t+1))+ G_{rj} cos(\delta_{rj}(t+1))] \\
\label{Pe_dis}
& Qe_r(t+1)=Q_{Lr}+V_r(t+1) \sum_{j=1}^{N} V_j(t+1)\nonumber \\ 
& \hspace{1.5cm} [G_{rj} sin(\delta_{rj}(t+1)) - B_{rj} cos(\delta_{rj}(t+1))]\\  \label{Qe_dis}
& V_m(t+1)=V_m^0 \\
& V_N(t+1)=V_N^0 \\
& \delta_N(t+1)=\delta_N^0.
\end{align}

where $j$ and $r$ represent bus numbers (all bus types), $N$ represents the total number of buses, and $N^{th}$ bus is the slack bus,  $m$ represents generator buses,  $Pm$ is the mechanical input power, $\omega$ is the rotor angular frequency, $\omega_s$ is the rotor synchronous angular frequency, $M=\frac{2H}{\omega_s}$, and $H$ is the inertia constant  \cite{crow2015computational}. $Pe$ and $Qe$ are real and reactive power generations, respectively, $P_L$ is active power load, $Q_L$ is reactive power load, $G$ is the real part of admittance matrix, $B$ is the imaginary part of admittance matrix, $V$ is the bus voltage magnitude and $\delta$ is the bus voltage angle. $V^0$ and $\delta^0$ are constants.  $\delta_{rj}=\delta_r-\delta_j$ in (7) and (8).

\vspace{10pt}
\section{Numerical Simulation}
\begin{figure}[b!]
\vspace{-5pt}
\centerline{\includegraphics[width=8cm]{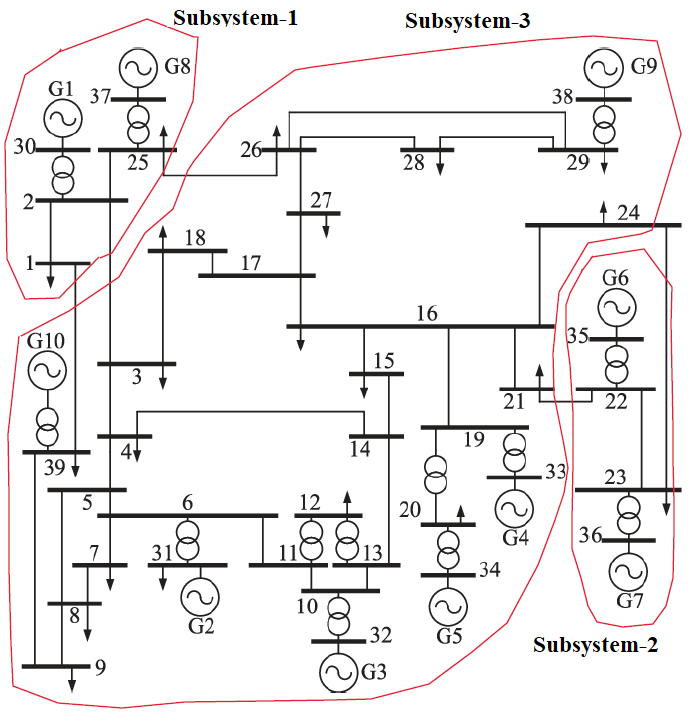}}
\caption{New England 10-generator 39-bus system~\protect\cite{pai2012energy}.}
\label{39bus}
\end{figure}

Numerical simulations are carried out using the New England 10-generator 39-bus system shown in Fig.\ref{39bus}. The system data are obtained from \cite{machowski2020power}. Newton-Raphson method is employed to solve the resulting nonlinear equations in the DI, WR and WRW algorithms. The power flow solution is used as the initial solution for the algebraic variables. At the beginning the system is assumed operating at synchronous speed. 
The simulation is carried out using a Windows machine with Intel (R) Core (TM) i7-8550U CPU@1.80 GHz 1.99 GHz processor, 8 GB of  RAM, and 64-bit operating system.

\subsection{Performance of Waveform Relaxation}
First, the test system is divided into three subsystems as shown in Fig. \ref{39bus}. Several simulations are run for the 20s (step size of 0.05s) with a series of disturbances. The first disturbance occurs at 0.2s, when the load on bus 29 disconnects and reconnects at 0.4s; then at 7.2s, the load on bus 25 is doubled for 0.2s; and at the 13.2s, the load at bus~23 is disconnected for 0.2s.   As each subsystem could be run in parallel in WR, the sum of the largest time taken by any subsystem in a particular WR iteration over all the required iterations is taken as the total time taken by the WR. 

\begin{figure}[h!]
\vspace{-15pt}
\centerline{\includegraphics[width=8cm]{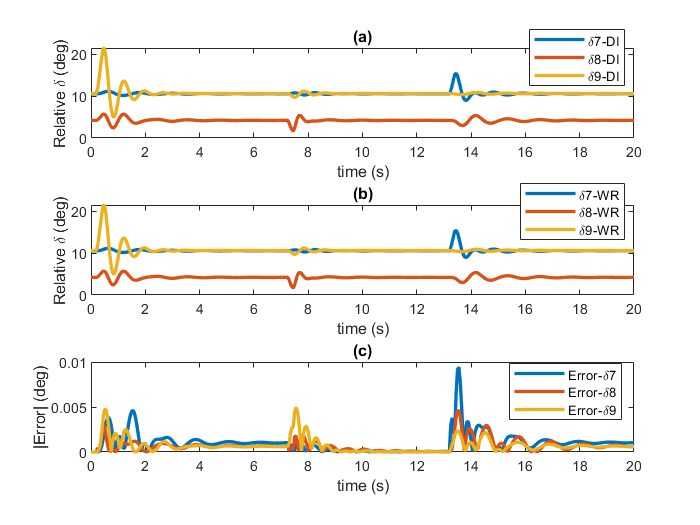}}
\vspace{-15pt}
\caption{Relative rotor angles: (a) DI, (b) WR, and (c) absolute error.}
\label{delta_diwr}
\end{figure}

\begin{figure}[h!]
\vspace{-15pt}
\centerline{\includegraphics[width=8cm]{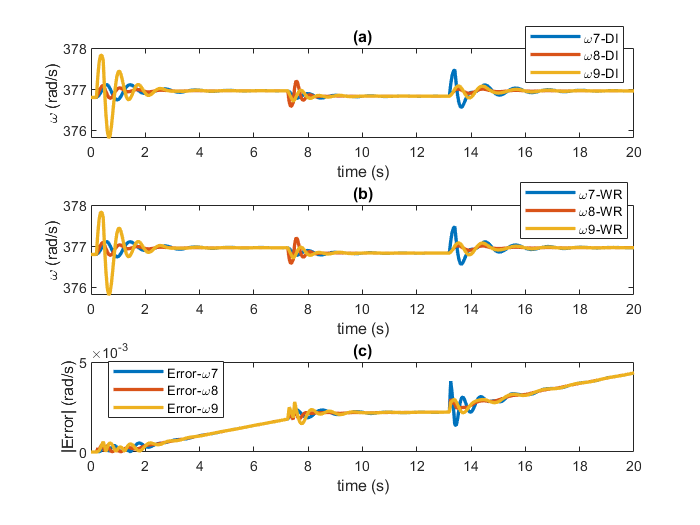}}
\vspace{-15pt}
\caption{Rotor frequencies: (a) DI, (b) WR, and (c) absolute error}
\label{omega_diwr}
\end{figure}

\begin{figure}[h!]
\vspace{-15pt}
\centerline{\includegraphics[width=8cm]{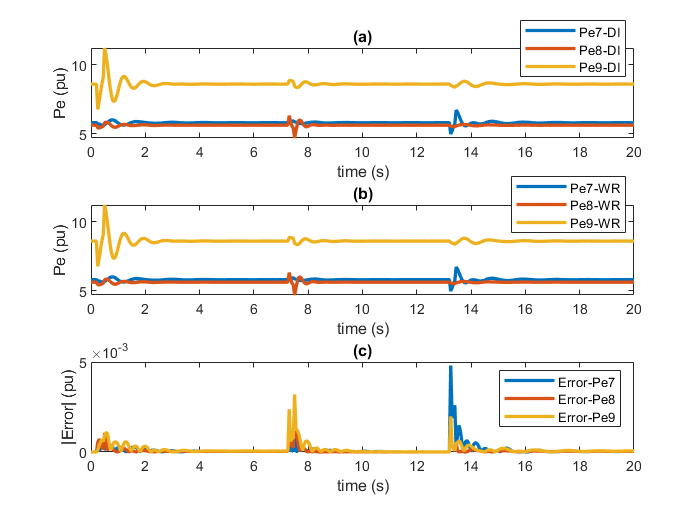}}
\vspace{-15pt}
\caption{Active power generations: (a) DI, (b) WR, and (c) absolute error}
\label{Pe_diwr}
\end{figure}

Comparisons of relative rotor angles (with respect to reference angle $\delta_N^0$), rotor angular frequencies, and active power generations obtained from WR for three generators (7, 8, and 9) are shown in Fig. \ref{delta_diwr}, Fig.~\ref{omega_diwr}, and Fig. \ref{Pe_diwr}, respectively, along with the solutions obtained from DI method.  DI and WR give fairly similar dynamic responses as the absolute errors are in the order of $10^{-3}$.  Although both DI and WR give very similar results, the solution time is 1.053s and 97.42s for DI and WR, respectively. Though, for the small-scale system as used in this work, DI seems promising, for large networks WR can outperform DI \cite{sangiovanni1985waveform}. 
 
\subsection{Performance of Waveform Relaxation with Windowing}
\begin{figure}[h!]
\vspace{-10pt}
\centerline{\includegraphics[width=8cm]{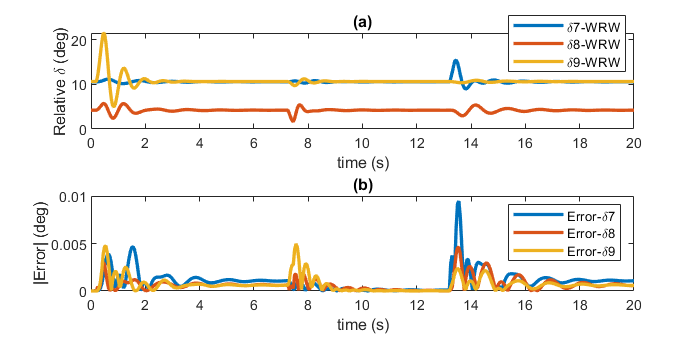}}
\vspace{-10pt}
\caption{Relative rotor angles: (a) WRW, and (b) error between DI and WRW.}
\label{delta_diwrw}
\end{figure}

\begin{figure}[h!]
\vspace{-10pt}
\centerline{\includegraphics[width=8cm]{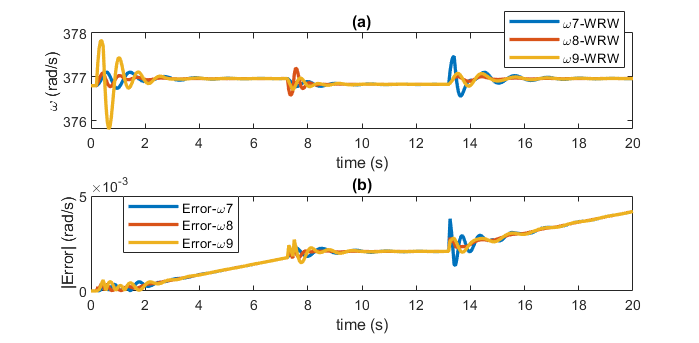}}
\vspace{-10pt}
\caption{Rotor frequencies: (a) WRW, and (b) error between DI and WRW.}
\label{omega_diwrw}
\end{figure}

\begin{figure}[t!]
\vspace{-15pt}
\centerline{\includegraphics[width=8cm]{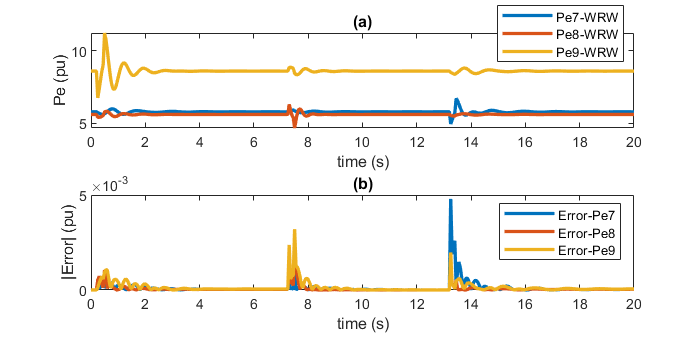}}
\vspace{-10pt}
\caption{Active power: (a) WRW, and (b) error between DI and WRW.}
\vspace{-10pt}
\label{Pe_diwrw}
\end{figure}

Simulations in Section IV. A are repeated using WRW. Total simulation time is divided into several windows where each window has an identical length equal to the step size of 0.05s, resulting in 400 windows for a 20s simulation. The solve time taken by the WRW is computed same as the WR but over all the windows.   Fig.~\ref{delta_diwrw}, Fig.~\ref{omega_diwrw}, and  Fig.~\ref{Pe_diwrw} respectively show the dynamics of the relative rotor angles, rotor frequencies, and active power generations obtained from WRW, and the corresponding errors compared with solution obtained from the DI method in Section IV.A.  The solution time taken by the WRW is
14.77s. Thus, the WRW is faster compared to WR. Moreover, unlike
the WR, the solution time (14.77s) for WRW is less than the total
simulation duration (the 20s); thus, WRW is promising for real-time performance \cite{jalili2010acceleration}.

{\color{black} The percent error on the solutions obtained from WR and WRW with respect to DI is also computed. The minimum, maximum, and average percent errors for select system variables are shown in Table~\ref{percent_error}. It is observed that the average percent errors for each of the variables are very small, which demonstrates the  accuracy of WR and WRW methods.}

\begin{table}[]
\vspace{-15pt}
\begin{center}
\caption{Percent error on select system variables}
 \resizebox{\columnwidth}{!}{\begin{tabular}{|c|c|c|c|c|c|c|}
\hline
\multirow{2}{*}{Variables} & \multicolumn{3}{c|}{WR}      & \multicolumn{3}{c|}{WRW}     \\ \cline{2-7} 
                           & Min.      & Max.   & Average  & Min.     & Max.    & Average  \\ \hline
$\delta_7$                     & -0.0024  & 0.8552 & 0.3409   & -0.0025  & 0.8132 & 0.3244   \\ \hline
$\delta_8$                     & -0.0019  & 0.8751 & 0.3506   & -0.0020  & 0.8322 & 0.3337   \\ \hline
$\delta_9$                    & -0.0029  & 0.8551 & 0.3411   & -0.0030  & 0.8132 & 0.3246   \\ \hline
$\omega_7$                    & -0.00009 & 0.0012 & 0.00056  & -0.00009 & 0.0011 & 0.00053  \\ \hline
$\omega_8$                      & -0.00011 & 0.0012 & 0.00056  & -0.00011 & 0.0011 & 0.00053  \\ \hline
$\omega_9$                     & -0.00015 & 0.0012 & 0.00056  & -0.00015 & 0.0011 & 0.00053  \\ \hline
$Pe_7$                        & -0.0967  & 0.0453 & -0.00053 & -0.0001  & 0.0453 & -0.0005  \\ \hline
$Pe_8$                        & -0.0496  & 0.0224 & -0.00051 & -0.0495  & 0.0224 & -0.00049 \\ \hline
$Pe_9$                        & -0.0379  & 0.0150 & -0.00043 & -0.0378  & 0.0151 & -0.00041 \\ \hline
\end{tabular}
}
\label{percent_error}
\vspace{-20pt}
\end{center}
\end{table}

\vspace{-5pt}
\subsection{Impact of the Number of subsystems}
The performance of WR and WRW depends on how the network is split into the subsystems. To investigate this, case studies are carried with multiple numbers of subsystems. Solve time in WR and WRW with a different number of subsystems is shown in Table~\ref{time4subsys}. It is observed that the solve times using WR and WRW are generally less for a higher number of subsystems, however, the solve time does not always decrease with an increased number of subsystems. One possible reason is that the network is arbitrarily divided into subsystems and, hence, the number of variables (and set of equations) are not evenly distributed over the subsystems. This leads to the variable size of the Jacobian matrices, which can impact the solve time. However, the solution time using the WRW method is always lower than the solution time using the WR method.

\begin{table}[t!]
\begin{center}
\caption{Solve time based on different subsystems}
\begin{tabular}{|c|l|c|c|}
\hline
\multirow{3}{*}{\textbf{\begin{tabular}[c]{@{}c@{}}No. of\\ Sub-\\ systems\end{tabular}}} & \multicolumn{1}{c|}{\multirow{3}{*}{\textbf{Subsystem Nodes}}}                                                                                                                                              & \multicolumn{2}{c|}{\textbf{Solve Time, s}}              \\ \cline{3-4} 
                                                                                          & \multicolumn{1}{c|}{}                                                                                                                                                                                       & \multirow{2}{*}{\textbf{WR}} & \multirow{2}{*}{\textbf{WRW}} \\
                                                                                          & \multicolumn{1}{c|}{}                                                                                                                                                                                       &                              &                               \\ \hline
2                                                                                         & \begin{tabular}[c]{@{}l@{}}SS1: 1,2,9,25,30,37\\ SS2: Rest of the nodes\end{tabular}                                                                                                                        & 133.35                       & 14.77                         \\ \hline
3                                                                                         & \begin{tabular}[c]{@{}l@{}}SS1: 1,2,25,30,37\\ SS2: 22,23,35,36\\ SS3: Rest of the nodes\end{tabular}                                                                                                       & 97.42                        & 11.54                         \\ \hline
4                                                                                         & \begin{tabular}[c]{@{}l@{}}SS1: 1,2,25,30,37,39\\ SS2: 19,20,33,34\\ SS3: 21,22,23,24,35,36\\ SS4: Rest of the nodes\end{tabular}                                                                           & 85.86                        & 15.25                         \\ \hline
5                                                                                         & \begin{tabular}[c]{@{}l@{}}SS1: 38,29,28,26,27,17,15,14,16,24\\ SS2: 39,1,2,30,25,37\\ SS3: 18,3,4,5,6,7,8,9,31,11,12,13,10,32\\ SS4: 19,20,33,34\\ SS5: Rest of the nodes\end{tabular}                     & 51.70                        & 11.11                         \\ \hline
6                                                                                         & \begin{tabular}[c]{@{}l@{}}SS1: 38,29,28,26,27,17\\ SS2: 37,25,2,30,1,39,3,18\\ SS3: 4,5,6,7,8,9,31,11,12,10,32,13\\ SS4: 19,20,33,34\\ SS5: 23,24,36\\ SS6: Rest of the nodes\end{tabular}                 & 73.81                        & 13.18                         \\ \hline
7                                                                                         & \begin{tabular}[c]{@{}l@{}}SS1: 1,2,25,30,37,39 \\ SS2: 38,29,28,26,27 \\ SS3: 36,23,24,16,21,22,35\\ SS4: 19,20,33,34 \\ SS5: 11,12,13,10,32 \\ SS6: 6,31,4,5,7,8,9 \\ SS7: Rest of the nodes\end{tabular} & 55.84                        & 6.86                          \\ \hline
\end{tabular}
\label{time4subsys}
\end{center}
\vspace{-20pt}
\end{table}

\subsection{Performance with Different Simulation Periods}
\begin{figure}[htbp]
\vspace{-10pt}
\centerline{\includegraphics[width=7cm]{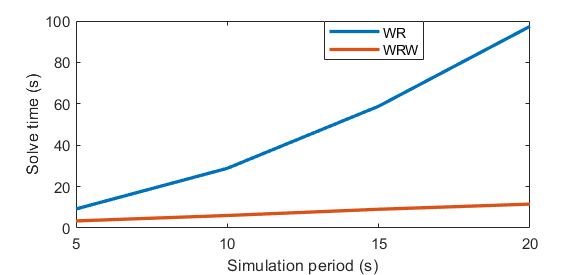}}
\vspace{-5pt}
\caption{Solve time  based on different simulation periods.}
\label{solution_time}
\end{figure}

Case studies are also run to test the impact of the simulation period on the solve time.  Fig. 
\ref{solution_time} shows the solve time of WR increasing non-linearly with the simulation period. In the WRW method, since the window size is the same as the step size, WRW solve time varies almost linearly with the length of the simulation period. It is worth to note that the average time taken by WRW to solve one window (of length 0.05s) is 0.031s, which is the desired attribute of any dynamical simulation method to implement on real-time simulators.
\vspace{-10pt}
\section{Conclusion and future work}
This paper has presented waveform relaxation and waveform relaxation with windowing as the alternatives to the direct integration methods for solving dynamical models of power systems. As waveform relaxation is parallel in nature, distributed computation is possible and could benefit large power systems when the direct integration method would not be faster in terms of solve time. However, from a real-time implementation point of view, waveform relaxation lacked the necessary solution speed.  The simulation using waveform relaxation with windowing technique showed promising results for faster dynamical simulation of power systems that could be implemented in real-time simulators. To further speed up the computation {\color{black} and check the scalability of the methods},  our future work will focus on optimal partitioning techniques to create the subsystems,  and time parallelization of waveform relaxation with the windowing method {\color{black}for larger networks}. 

\vspace{-6pt}
\bibliography{biblio_ISGT2021.bib}
\bibliographystyle{IEEEtran}

\end{document}